\documentclass[a4paper]{article}

\usepackage{subfigure}
\usepackage{epstopdf}
\usepackage{authblk} 
\usepackage{amsmath,amssymb}
\usepackage{graphics}

\begin{document}

\title{
Characterization, 1064~nm photon signals and 
background events of a tungsten TES detector 
for the ALPS experiment}

\author{
J. Dreyling-Eschweiler$^{\rm a}$$^{\ast}$\thanks{$^\ast$Corresponding author. Email: jan.dreyling-eschweiler@desy.de\vspace{6pt}},
N.~Bastidon$^{\rm b}$,
B.~D\"obrich$^{\rm a}$,
D.~Horns$^{\rm b}$,
F.~Januschek$^{\rm a}$,
and A.~Lindner$^{\rm a}$.
\\
\vspace{6pt}
$^{a}${\em{Deutsches Elektronen-Synchrotron (DESY), Hamburg, Germany}}; \\
$^{b}${\em{Institut f\"{u}r Experimentalphysik, Universit\"{a}t Hamburg, Germany}}
}

\date{February 27, 2015 (DESY 14-193) \\ 
Accepted by Journal of Modern Optics \\
(DOI: 10.1080/09500340.2015.1021723)}

\maketitle

\begin{abstract}
The high efficiency, low-background, and single-photon detection with transition-edge sensors (TES) is making this type of detector
attractive in widely different types of application.
In this paper, we present first characterizations of a TES to be used in the Any Light Particle Search (ALPS) experiment searching for new fundamental ultra-light particles.
Firstly, we describe the setup and the main components of the ALPS TES detector (TES, millikelvin-cryostat and SQUID read-out) and their performances.
Secondly, we explain a dedicated analysis method for single-photon spectroscopy and rejection of non-photon background.
Finally, we report on results from extensive background measurements. 
Considering an event-selection, optimized for a wavelength of 1064~nm,  we achieved a background suppression of $\sim 10^{-3}$ with a $\sim 50$~\% efficiency for photons passing the selection. 
The resulting overall efficiency was 23~\% with a dark count rate of $8.6\cdot10^{-3}$~s$^{-1}$.
We observed that pile-up events of thermal photons are the main background component.
\end{abstract}

\newpage

\tableofcontents

\section{Introduction}
\label{sec:intro}

Over the last few years, the development of optical transition-edge sensors used as single-photon detectors has resulted in a near-unity detection efficiency \cite{lita2008, fukuda2011}. 
Besides the application in Quantum Optics experiments \cite[e.g.]{giustina2013}, there is an application in the Any Light Particle Search (ALPS) experiment,  
which we describe in this paper.

{The ALPS experiment is a light-shining-through-a-wall experiment \cite{Ehret2010149,alpsII_tdr}:
Photons are stored in front of an optical barrier, behind which, a photon detector is placed. 
Standard Model physics does not allow for light passing the wall with a measurable rate.
However, photons could convert into low-mass particles beyond the Standard Model, which would interact very weakly with the barrier. In this way they could effectively traverse the wall and reconvert into a photon behind the obstacle.}
Such new particles are weakly interacting sub-eV particles (WISPs) \cite{jaeckel2010}, including the axion which has been proposed to solve the strong CP-problem \cite{peccei1977,wilczek1977,weinberg1977}, are well motivated by theoretical considerations, and could explain several astronomical and cosmological observations \cite[e.g.]{hewett2012, essig2013} 
including non-baryonic Dark Matter \cite[e.g]{arias2012}. 

{Using an enhanced setup including an increased magnetic length, optical cavities to amplify the signal and a low-background single-photon detector, 
the sensitivity of the second generation of ALPS \cite{alpsII_tdr} will be improved by three orders of magnitude with respect to the previous generation of this type of experiment \cite{Ehret2010149}.}
In order to reach the sensitivity required, the single-photon detector should have an overall detection efficiency $>70~\%$ for
photons with a wavelength of 1064~nm {(and an energy of 1.165~eV)}, while registering a background not exceeding a few mHz \cite{alpsII_tdr}.
We found that a transition-edge sensor (TES) is the most promising detection technique \cite{seggern2013}. 

A TES acts as a sensitive microcalorimeter operated in the superconducting transition \cite{irwin2005}. 
Here, small temperature changes, like the absorption of an infrared photon ($\sim 1$~eV), cause measurable changes of the electrical resistance. Due to its calorimetric nature -- events with higher energy result in a higher resistance change -- 
a TES acts as a non-dispersive single-photon spectrometer with a coarse energy resolution. 
In the range from optical to near-infrared photons, it is about $\Delta E/E \lesssim 10$~\% \cite{lita2008, fukuda2011}.

In this paper, we describe the ALPS TES detector and its main components (Sec.~\ref{sec:alps_tes_detector}). 
We report on results of stability and linearity measurements (Sec.~\ref{sec:perform}).
We explain a newly developed method to analyze the pulse shape of events and to calibrate the TES (Sec.~\ref{sec:signal}).
We report on our results of long-term background measurements and of resulting dark count rates (Sec.~\ref{sec:background}).
Finally, we summarize our characterization of the ALPS TES detector (Sec.~\ref{sec:concl}).

\section{Components}
\label{sec:alps_tes_detector}

The first efforts to install and operate a TES detector for the ALPS experiment commenced in 2011 \cite{de2013}.
Since the end of 2013, a completed TES detector system has been operated routinely in the ALPS laboratory, namely the ALPS TES detector (Fig.~\ref{fig:adr}).
We briefly explain the three main components and their performance. A detailed description is found in \cite{jde2014}.

\subsection{Sensor: Fiber-coupled NIST TES}
\label{sec:alps_tes_detector:tes}

\begin{figure}
\begin{center}
\begin{minipage}{130mm}
  \begin{minipage}{60mm}
      \subfigure[sketch of ALPS TES detector]{
	\label{fig:adr:sketch} 
        \resizebox*{6cm}{!}{
          \includegraphics{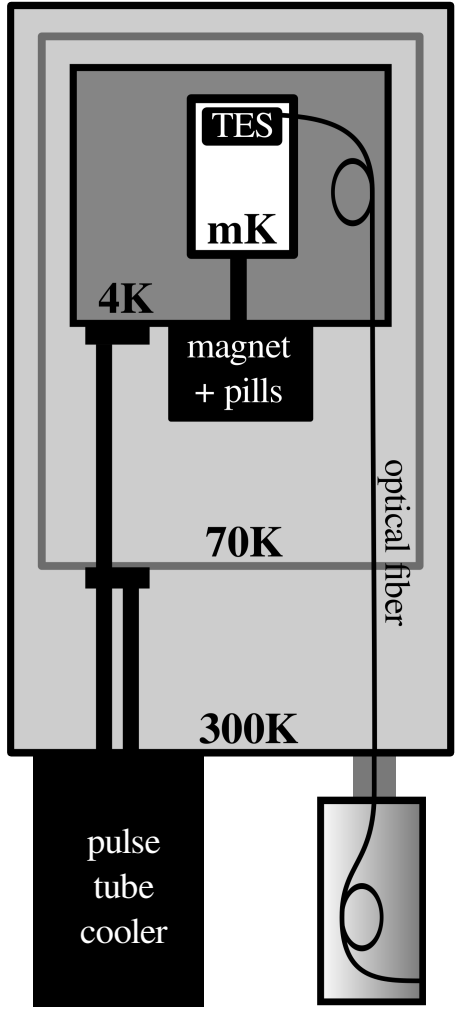}
        }
      }
  \end{minipage}
  \hspace{6pt}
  \begin{minipage}{60mm}
    \subfigure[sensor module with two channels]{
	\label{fig:adr:tes} 
    \resizebox*{6cm}{!}{
        \includegraphics{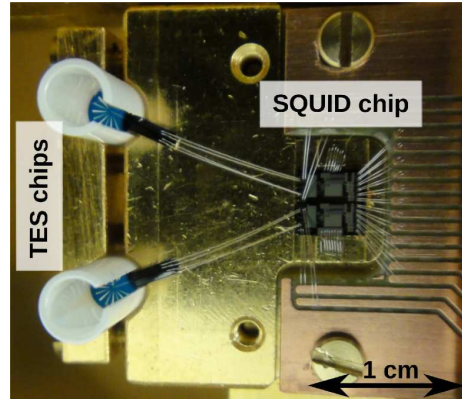}
      }
    }
      \\
      \subfigure[opened ADR cryostat]{
	\label{fig:adr:adr} 
    \resizebox*{6cm}{!}{
        \includegraphics{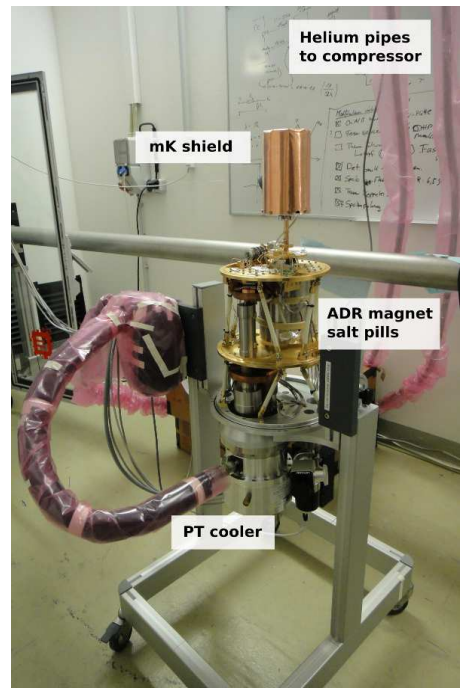}
      }
    }
  \end{minipage}
\caption{{Sketch and} components of the ALPS TES detector: 
{(a) shows a sketch of the ALPS TES detector incorporating the different cooling stages of the ADR, the TES device and the optical fiber which is guided outside of the cryostat.}
(b) shows a sensor module with two channels. The round NIST TES chips can be found on the left surrounded by a fiber sleeve. On the right, the PTB SQUID chip incorporating two channels is located. Aluminum wire bonds are used for electrical connection.
(c) shows the opened ADR cryostat with the different cooling stages. The pulse-tube cooler cools the upper copper plate. On this plate, the salt pills unit with the magnet is mounted. 
The superconducting devices are located on the top, inside the copper mK-shield. 
In the back, a part of the 20~m long helium pipes connecting the PT cooler to a compressor can be seen.
}
\label{fig:adr} 
\end{minipage}

\end{center}
\end{figure}

We are using a TES provided by NIST\footnote{National Institute of Standards and Technology.} \cite{lita2010}. 
The sensitive material is a 20~nm thin tungsten film which has a sensitive area of 25~$\mu$m $\times$ 25~$\mu$m. 
The film is incorporated in an optical stack enhancing the detection efficiency of 1064~nm photons.
The TES chip is surrounded by a standard fiber connector sleeve (Fig.~\ref{fig:adr:tes}) which allows to connect a single mode fiber \cite{miller2011}. 
This method is cryogenically compatible and is estimated to show a low optical loss of $< 0.1$~\% in between the single mode fiber diameter ($<10$~$\mu$m) and the TES area.
For such fiber-coupled TES, a detection efficiency of {higher than 95~\% was reached \cite{lita2010}.}
Knowing that this is possible in principle, our main aim in these first studies  was to investigate backgrounds relevant for the ALPS experiment. 
For the measurements presented in this paper, we have used standard SMF28 fibers with an anti-reflective coating for 1064~nm.%
\footnote{The coating reduces the standard 4~\%-loss at a fiber end to $< 1$~\%.}

In the commissioning phase of the detector system, we determined further important parameters of the used TES chip\footnote{In the current setup, we operate two TES chips from NIST: channel A and B. In this paper, we only refer to channel A which is characterized in more detail. For channel~B, we have found comparable results.} \cite{jde2014}.
We have determined a superconducting transition of $T_{\rm c} = 140$~mK with a width of $\Delta T_{\rm c} =$ 1-2~mK.
All following results are referring to the same chosen working point:
\begin{itemize}
  \item a bath temperature of $T_{\rm b} = 80$~mK and 
  \item a TES set point, $R_0$, within its superconducting transition of 30~\% of the normal resistance, $R_{\rm N} = 13.8$~$\Omega$, which is reached by applying a {constant bias voltage across the TES}.
\end{itemize}
Thus, the TES has been operated in a strong negative electro-thermal feedback 
{(ETF): Due to the constant bias voltage, the Joule power $P_{\rm J}= V^2/R$ at the TES corresponds to the power flow to the bath. If the TES heats up, the resistance increases, and the Joule heating is decreased. Thus, the cooling of the TES is faster. Furthermore, the ETF}
stabilizes the operation as well as simplifies theoretical analysis of the TES \cite{irwin2005}.

The heat capacity, $C = 0.6$~fJ/K, and the thermal conductance, $G = 15$~pW/K, are calculated according to experience with similar optical tungsten-based TES \cite{lita2005}. 
This results in a thermal time constant $\tau = C/G \approx 38$~$\mu$s.
{Due to the ETF, the time constant of a TES is effectively reduced to $\tau_{\rm eff }= \frac{\tau}{1+\alpha/n}$, where $\alpha = T/R \cdot {\partial R}/{\partial T}$ which is the steepness of the resistance-temperature dependence at the set point $R_0$ and $n$ expresses the thermal impedance between the bath substrate and the superconducting film which is 5 or 6 for tungsten \cite{irwin1995etf}. 
Using an averaged pulse, we determine $\tau_\mathrm{eff}=1.53~\mu\mathrm{s}$, such that $\alpha/n=24$ and therefore find $\alpha$ to be between $120$ ($n=5$) and $144$ ($n=6$).
}
{Furthermore, we determined an energy loss of $\sim 10$~\% when comparing the measured pulse to the nominal photon energy \cite{irwin2005}. This is assumed to be explained by high-energetic phonons escaping from the tungsten film during the absorption process, so that only a fraction of the photon energy is measured, as described in \cite{cabrera1998}.}
We also {estimated} the TES saturation {-- when the TES  becomes temporarily normal resistive due to an energy input --} which {would} correspond to a 330~nm single-photon event or of more than three simultaneous 1064~nm single photons.
The effect of saturation becomes already evident in the linearity measurement for 405~nm photons {(Sec.~\ref{sec:perform:linear} and Fig.~\ref{fig:linearity})}. 
Finally, a measured average photon pulse has been proven to be in accordance with the small-signal TES theory \cite{irwin2005} by using the determined TES parameters.

\subsection{Millikelvin-temperature: Adiabatic Demagnetization Refrigerator (ADR)}
\label{sec:alps_tes_detector:mk}

To reach mK-temperatures, we operate a cryogen-free cryostat, namely an Adiabatic Demagnetization Refrigerator (ADR). 
  The ADR system from Entropy GmbH is constructed following a top-down principle where mK-temperatures are reached at the innermost stage (Figs.~\ref{fig:adr:sketch} and \ref{fig:adr:adr}). 
A pulse tube cooler with a closed $^{4}$He gas cycle provides the precooling below 4~K.
The adiabatic demagnetization cooling \cite{white2002} is realized by a salt pills unit acting as a paramagnetic spin-system, which is surrounded by a superconducting magnet. 
When the field of the magnet is ramped up to 6~T, the spin-system is aligned.
After the pills unit is thermalized, it is thermally disconnected from the precooling stage.
By ramping down the magnetic field, a low-temperature below 100~mK can be reached at the innermost stage.

The evacuation and precooling take $\sim25$~h depending on the installed cold mass \cite{jde2014}. 
Charging or recharging the pills unit takes $\sim90$~min depending mainly on the performance of the adjustable heat switch driven by a piezo motor. 
Regulating a small remnant magnetic field allows to stabilize the inner temperature to $80$~mK $\pm$ $25$~$\mu$K (rms) for $\gtrsim 20$~h. A recharge with a subsequent mK-regulation is named recharge-cycle in the following.  
We could operate the ALPS TES detector for instance in a 24~h cycle consisting of 20~h measurement time and 4~h dead time due to the recharge and sensor adjustments.

\subsection{Read-out: Low-noise SQUID}
\label{sec:alps_tes_detector:read_out}

The TES chip has been integrated in an electrical circuit for two reasons. 
Firstly, this allows to bias accurately the device in its superconducting transition. 
Secondly, it enables measurements of the TES resistance variations. Therefore, the TES circuit is inductively coupled to a SQUID\footnote{Superconducting Quantum Interference Device.} sensor. 
We are using a 2-stage SQUID  (Fig.~\ref{fig:adr:tes}) from PTB\footnote{Physikalisch-Technische Bundesanstalt.} \cite{ptb_squids2007} which provides a low-noise current read-out of the TES circuit.
The SQUID device is stably working at mK-temperatures and has a noise level of approximately 2~pA/$\sqrt{\rm Hz}$. 

The TES and SQUID chips are mounted to the inner ADR stage (Figs.~\ref{fig:adr:sketch} and \ref{fig:adr:tes}). 
Superconducting or screened cables are used to connect the sensors to the electronics located outside the ADR cryostat.
Using an electronics from Magnicon GmbH, we can set bias currents for the TES and SQUID as well as read out the signal output. 
The read-out system has a bandwidth of $f_{3\rm dB} \approx 0.9$~MHz.
To record the signal output, we have used a commercial oscilloscope\footnote{Tektronix DPO7104C$^{\circledR}$.}. 
We set the sampling interval to $\Delta t= 50$~ns and the digitization step to 0.4~mV. 
For signal calibration (Sec.~\ref{sec:signal:calib}), we have recorded time series with 20 Msamples covering each a 1~s window.
For long-term background measurements (Sec.~\ref{sec:background}), we have recorded triggered single events with an extended read-out window going from 5~$\mu$s before to 15~$\mu$s after the trigger.
{We observe a dead time of the oscilloscope below 0.5~seconds when recording a single event. Concerning the long-term background measurements (Sec.~\ref{sec:background}), we adjust the trigger level to have a trigger rate below $0.1$~s$^{-1}$ and not to be affected by this dead time.}

\section{Performance tests}
\label{sec:perform}

\subsection{Stability}
The stability of the ALPS TES detector and thus its reliability is of special importance for the ALPS experiment, where long-term measurements are planned, looking for possible signal photons having a very low rate. 
Thus, the signal region needs to be stably defined without a necessity to readjust it in short time intervals.

We have investigated the stability of the TES output operated in the ADR with regards to 
\begin{itemize}
\item different times during one recharge-cycle, i.e. after recharging the pills unit when regulating with a remnant magnetic field, 
\item different recharge-cycles,  
\item different cool-downs (after allowing the system to thermalize to room temperature and cooling it down again) and 
\item human operator due to the adjustment method.
\end{itemize}
With regards to all these factors the system was found to be reasonably stable, i.e. the working point and the signal region do not differ considerably.

\begin{figure}
\begin{center}
\begin{minipage}{140mm}
  \subfigure[stability measurement]{
      \label{fig:stabr} 
    \resizebox*{7cm}{!}{\includegraphics{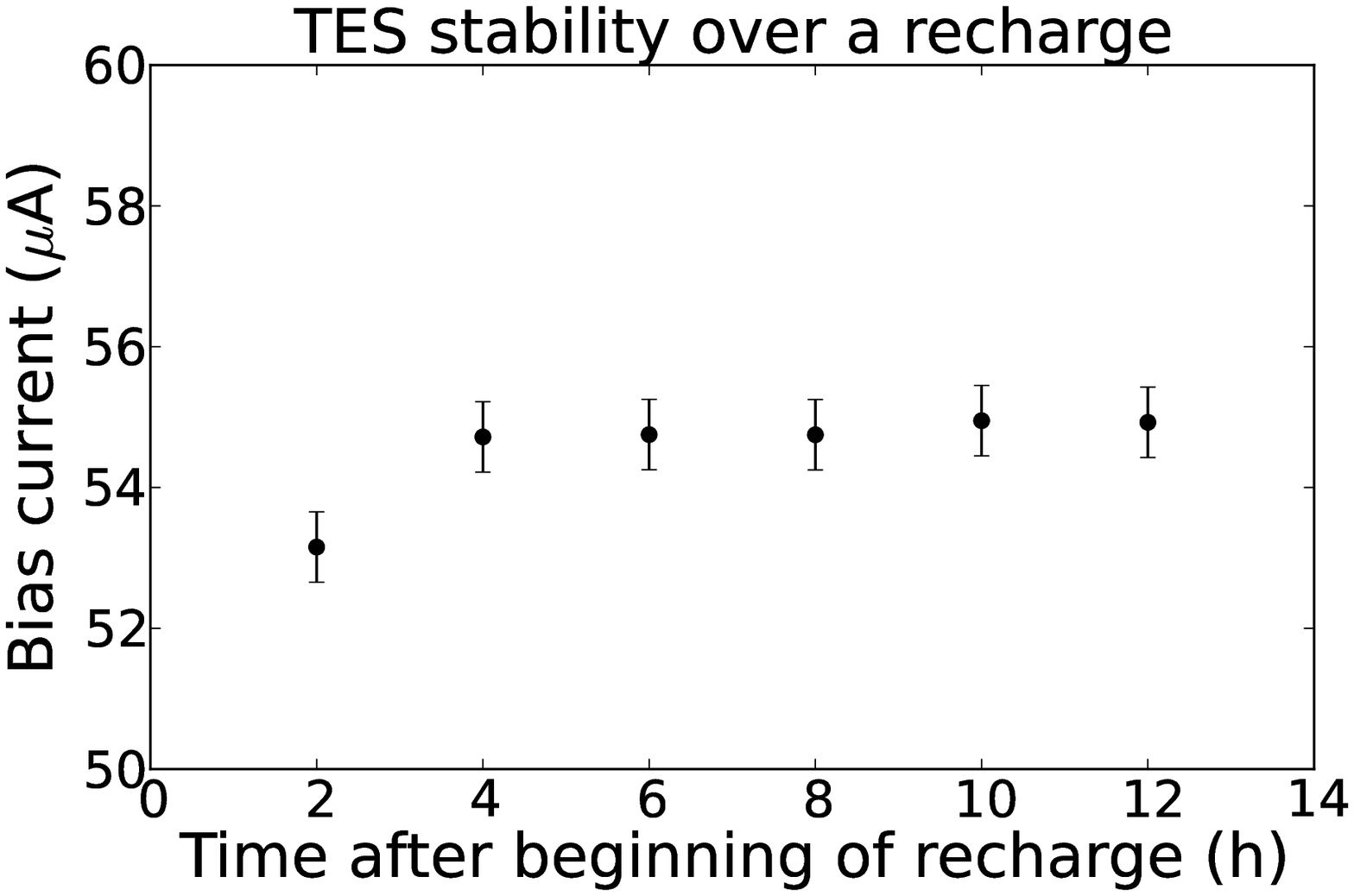}}}
  \hspace{6pt}
  \subfigure[linearity measurement]{
      \label{fig:linearity} 
      \resizebox*{7cm}{!}{\includegraphics{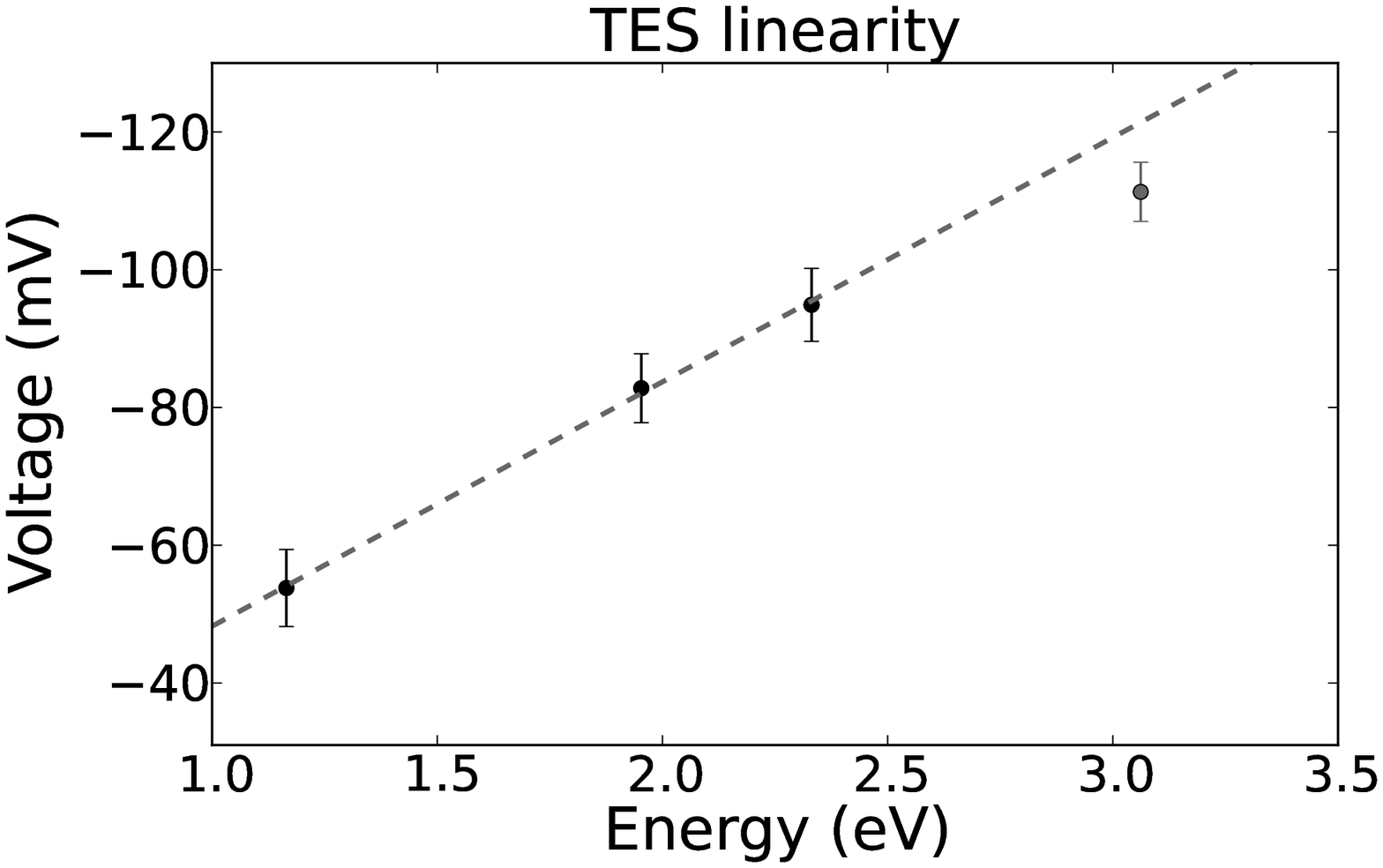}}}
\caption{
  (a) Stability: the TES bias current equivalent to $R_0 = 30$~\% $R_{\rm N}$ as a function of time after the beginning of a recharge.
 (b) Linearity:  
 average pulse height in units of voltage output as a function of photon energy for the TES. The error bars show the widths of the respective Gaussian distributions. {The dashed line is a fit to the first three points.}
 }
\end{minipage}
\end{center}
\end{figure}

Of special importance for the ALPS experiment is the stability within a recharge-cycle to reduce the overall uncertainties.
We were able to show that our working point, i.e. the TES bias current for the chosen $R_{0}$ (Sec.~\ref{sec:alps_tes_detector:tes}), is approximately stable during the course of a recharge. 
The values of the bias current only vary by about $<1.5$~$\mu$A in maximum (Fig.~\ref{fig:stabr}) which results in a $<3$~\%-change of the TES pulse height.

\subsection{Linearity} 
\label{sec:perform:linear}

We have carried out linearity measurements regarding the response of our TES detector setup to photons of different wavelengths. 
For that purpose, we used lasers of four different wavelengths in the optical and near-infrared.%
\footnote{Nominal wavelength of 405~nm, 532~nm, 635~nm and 1064~nm.}
For each wavelength, we set the laser rate with attenuators to reduce the event rate such that the pile-up rate was negligible. 
We recorded several pulse height distributions. 
All the laser sources resulted in approximately Gaussian distributions of the pulse peak height. 
The different signal regions were fitted with a Gaussian. 
The absolute widths of the Gaussians were approximately constant due to the constant sensor noise, while it results in relative widths of 4~\% (at 405~nm) to 10~\% (at 1064~nm) of the corresponding mean values. 
By plotting the mean values of the Gaussians as a function of photon energies (Fig.~\ref{fig:linearity}), it can be seen that the TES response is indeed approximately linear in the region relevant to the ALPS experiment, though results indicate a non-linear behavior for higher energies, which can be understood as a saturation of the detector (Sec.~\ref{sec:alps_tes_detector:tes}).

\section{Event analysis and signal region} 
\label{sec:signal}

An event is selected if the registered amplitude crosses the chosen (online or offline) trigger level ($TL$). 
The corresponding event window includes data points before and after the trigger point, $t=0$ (Sec.~\ref{sec:alps_tes_detector:read_out}). 
These data points are analyzed to extract different parameters. 
By using such parameters, events can be classified.

The first common approach is to consider the pulse height ($PH$) as the maximum level of an event window.
This allows to accumulate events in a pulse height distribution (PHD) and to define a one-dimensional signal region.
A second parameter is calculated by taking into account the pulse integral ($PI$).
Combining the two parameters in a two-dimensional distribution, 
it is possible to discriminate between single-photon events with different energies as well as, pile-up events or baseline shifts causing a trigger \cite{jde2014}. 
Finally, the available information of the pulse-shape can be extracted by fitting an averaged pulse profile to the read-out window. 
This pulse shape analysis (PSA) achieves the best background discrimination.

\subsection{Pulse shape analysis (PSA)}
\label{sec:signal:psa}

In this analysis, an expected single-photon pulse shape is fitted to an event. 
As the expected pulse shape, we use an averaged pulse of single 1064~nm photon signals.%
\footnote{The averaged pulse results from measurements including only events which are lying in the 3$\sigma$-region of the extended two-dimensional PHD. 
  Using the average pulse as the expected photon pulse shape, the limited read-out bandwidth of the read-out system is taken into account automatically. 
  Furthermore, the expected pulse is a self-consistent estimate of the overall system and does not depend on the first-order TES theory \cite{irwin2005}.}
The average pulse is shifted by an integer multiple $j$ of the sampling interval $\Delta t$ (Sec.~\ref{sec:alps_tes_detector:read_out}) before or after the trigger point. 
For each shift we calculate the best fit by scaling the average pulse shape with a dimensionless pulse scaling factor~$a$. 

Thus, the fit function $f_j$ is
$$f_j(t, a) = a \cdot {\rm avg}_j(t) \ ,$$
where ${\rm avg}_j$ is the average pulse shifted by the index $j$.
The best fit is determined by choosing $a$ such that $\chi^2_j$ is minimized. It is defined as 
$$\chi_j^2 = \sum_{i=1}^N \left(\frac{y_i - f_j(t_i, a)}{{\rm d}y_i}\right)^2 \ .$$ 
We consider $N$ data points of each event window and of the average pulse. 
$y_i$ is the $i$-th voltage entry of the event, and $f_j(t_i, a)$ the $i$-th voltage entry of the scaled average pulse. 
We determine the uncertainty ${\rm d}y_i$ by considering the standard deviation of the noise band $\sigma_{\rm noise}$. We fix ${\rm d}y_i = \sigma_{\rm noise} = 5.0\pm0.1$~mV for each data point.

\begin{figure}
\begin{center}
\begin{minipage}{140mm}
  \subfigure[single event analysis]{
      \label{fig:single} 
    \resizebox*{7cm}{!}{\includegraphics{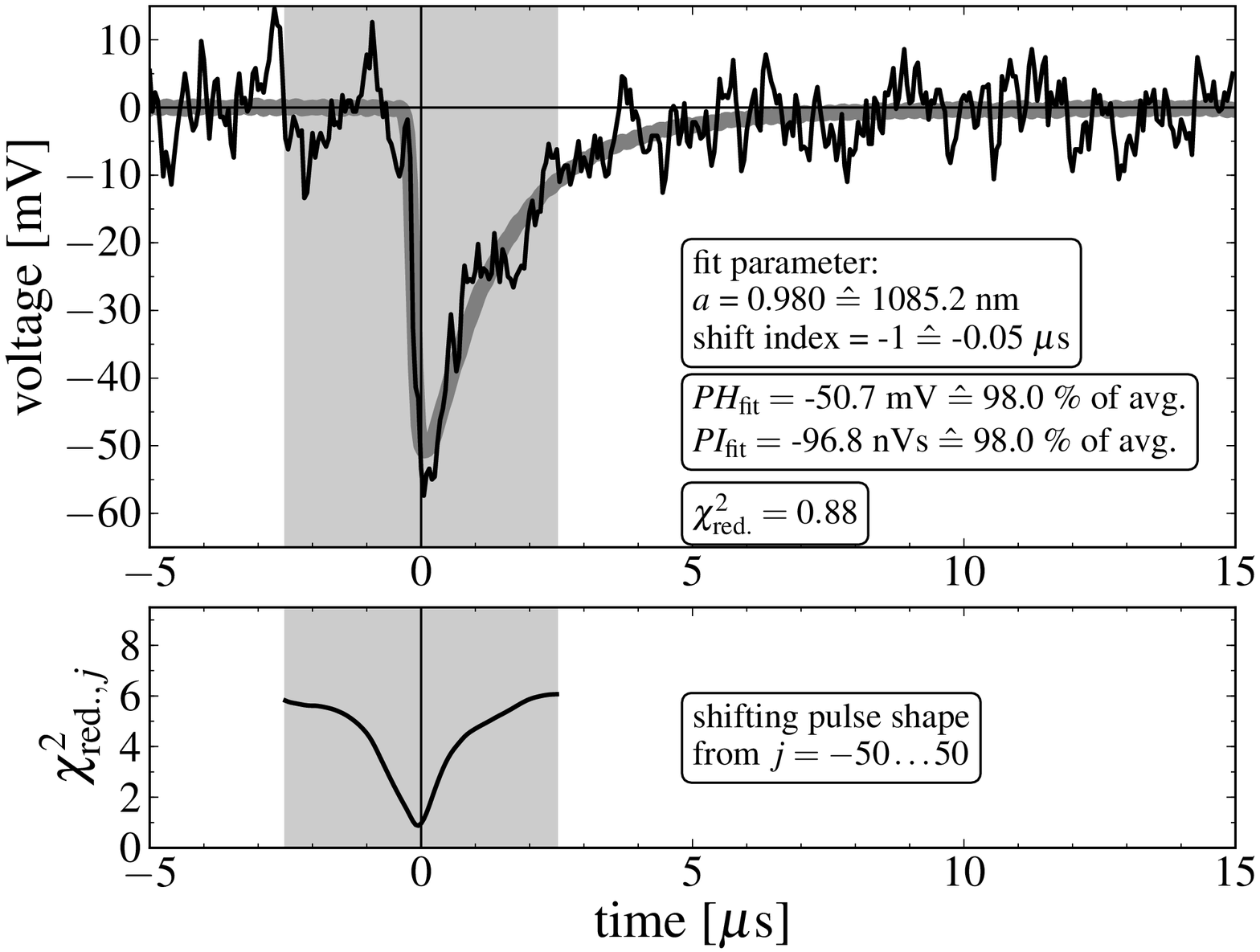}}}
  \hspace{6pt}
  \subfigure[1064~nm signal region]{
      \label{fig:signal} 
      \resizebox*{7cm}{!}{\includegraphics{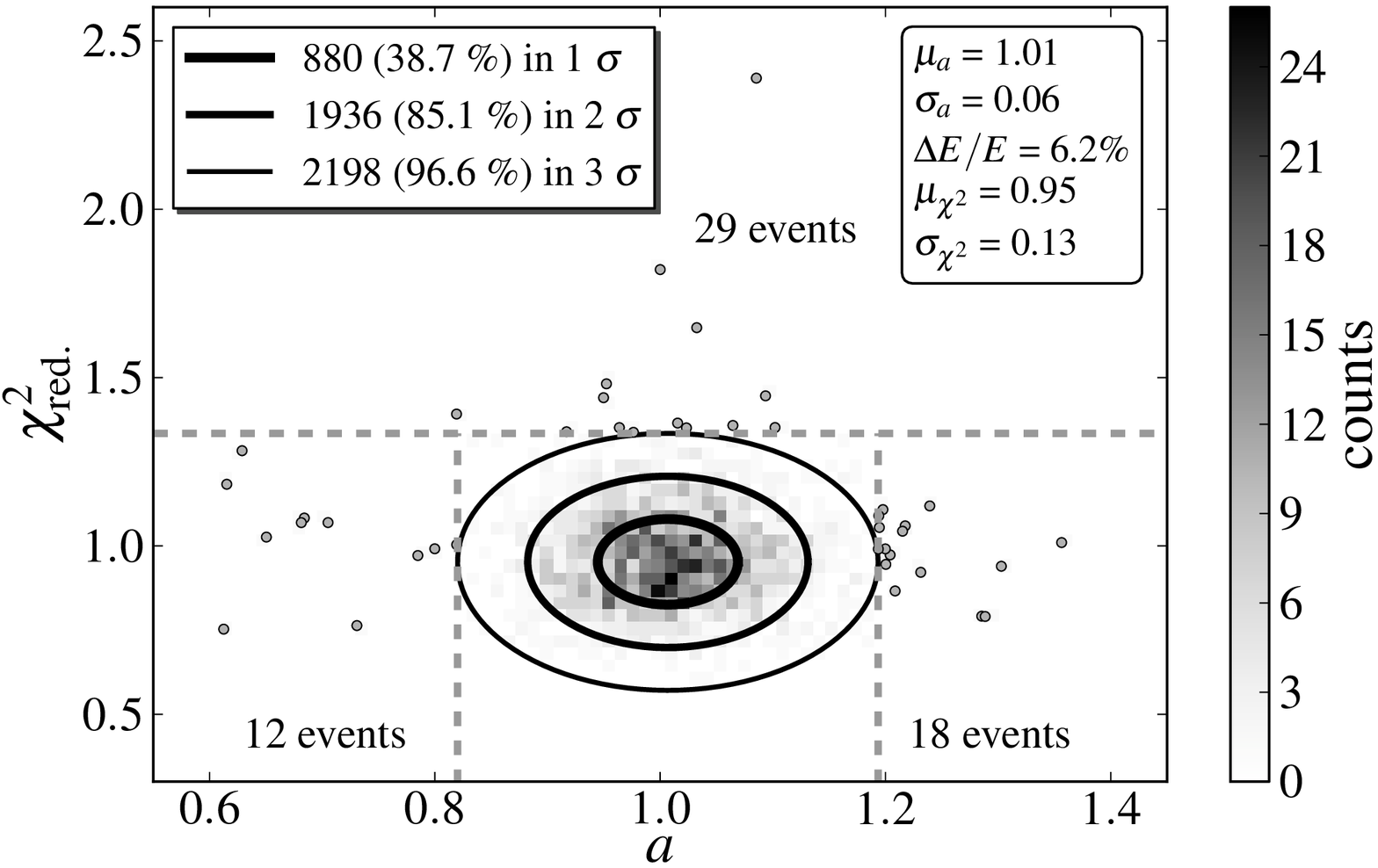}}}
\caption{Application of the pulse shape analysis (PSA): (a) shows the best fit to a single event. The thick gray line in the upper plot illustrates the average 1064~nm pulse. The grayish area illustrates the range of shifting the index, here $j= -50 \ldots 50$. The lower plot illustrates the significant minimum of chi-squared due to the best fit. 
  (b) shows the two-dimensional distribution of signal 1064~nm photons. This allows to define elliptic $\sigma$-regions. Using 3$\sigma$-cuts along both axes, it is found that the signal sample contains few impurities.
}
\end{minipage}

\end{center}
\end{figure}

Finally, we can determine the best shift $j$ by finding the minimal $\chi_j^2$. 
Thus, by analyzing an event with the PSA, we extract two useful quantities: 
the  $\min(\chi_j^2) \equiv \chi^2$ and the corresponding scaling factor $a$ of the average pulse.
A typical event with the best fit is illustrated in Fig.~\ref{fig:single}.

\subsection{Signal calibration and event classes}
\label{sec:signal:calib}

To calibrate the ALPS TES detector, we operate the TES with a fiber attached to a single-photon source. 
As a source we use an attenuated, continuous wave, 1064~nm laser.%
\footnote{The nominal wavelength of the used laser is 1063.9~nm. The corresponding energy deviation is much smaller than the energy resolution of the TES detector, $\Delta E/E < 10$~\%.}
The attenuation is adjusted to reach a single-photon rate of $10^3$ to $10^4$~s$^{-1}$.

We analyze 5 seconds of time series offline. Firstly, we extract single events by applying a trigger level which is 3$\sigma$ below the mean pulse height in a one-dimensional PHD. 
This excludes low-energetic events with smaller values of pulse heights.
Secondly, we apply the PSA method to each extracted event. In the following, we classify the events of the 5~s sample by considering their distribution in the $\chi_{\rm red.}^2$-$a$-plane (Fig.~\ref{fig:signal}):%
\footnote{Instead of the $\chi^2$, we consider the reduced chi-squared which is defined by $\chi_{\rm red.}^2 = \frac{\chi^2}{\nu}$, where $\nu$ is the degree of freedom.
In this example, it is $\nu = N - n = 400 - 2 = 398$, which includes all $N=400$ data points of an event window and $n=2$ due to two free fit parameters, the scaling factor $a$ and the shift index $j$.}
\begin{itemize}
      \item \textbf{Signal events} (events in the 3$\sigma$-region): 
	It is assumed that these events are single signal photons from the laser source. This is validated by analyzing a sample with laser off {(e.g. Fig.~\ref{fig:fiber})}.
	Signal events normally distribute around $\mu_a = 1.0$ and $\mu_{\chi^2_{\rm red.}} = 0.95$. 
By fitting a two-dimensional Gaussian distribution, we define 1-, 2- and 3$\sigma$ signal regions which are illustrated by the elliptic contours.

  \item \textbf{Low-energetic single photon-like events} ($\chi^2_{\rm red.}< \mu_{\chi^2_{\rm red.}} + 3 \sigma_{\chi^2_{\rm red.}} = 1.34$ and $a<\mu_a-3\sigma_a = 0.82$):
	Few events of this sample are presumably related to low-energetic photons between $\sim$1300-1950~nm. 
	If these events are due to single photons, they presumably originate from thermal background or {from outlying signal events. For this sample, we expect approximately 8 signal events to be located in this region  to be compared to the 12 events detected.} 

  \item  \textbf{High-energetic single photon-like events} ($\chi^2_{\rm red.}< \mu_{\chi^2_{\rm red.}} + 3 \sigma_{\chi^2_{\rm red.}} = 1.34$ and $a>\mu_a+3\sigma_a=1.20$):  
	Few events are presumably related to single high-energetic photons between  $\sim$600-900~nm. 
	If these events are due to  single photons, they could originate from stray light or 
	{from outlying signal events. For this sample, we expect approximately 8 signal events to be located in this region to be compared to the 18 events detected.}
  
  \item \textbf{Pile-up or background events} ($\chi^2_{\rm red.}> \mu_{\chi^2_{\rm red.}} + 3 \sigma_{\chi^2_{\rm red.}} = 1.34$): 
    {For large values of $\chi_{\rm red.}^2$, the contribution of outlying signal events is small (8 expected out of 29 detected events). 
  Most likely these additional events are} first-order pile-up events. 
A first order photon-like pile-up event is defined as two photon-like events that are detected within the same time window. 
   It is shown that such events are well discriminated by the PSA if the two constituents occur within a time window $\gtrsim 0.5~\mu$s (Sec.~\ref{sec:background:thermal}).
    Other background events can also result in a high $\chi^2_{\rm red.}$ (Sec.~\ref{sec:background:intrinsic}).  
\end{itemize}

Furthermore, we can determine an energy resolution for this sample related to 1064~nm photons:
  \begin{eqnarray}
    (\Delta E/E)_{1064~nm} =  |\sigma_{\rm a} / \mu_{\rm a}| \approx 6.2 {\rm ~\%}\ .
\label{eqn:erg_res:psa}
\end{eqnarray}
Finally, the scaling factor $a$ provides a calibration.
$a=1$ is calibrated to the pulse height of the average pulse.
Thus, $a$ directly determines the energy of a well fitted event. 
The calibration for photon energies and corresponding wavelengths is listed in Table~\ref{tab:fit_calibration}.

\begin{table}
\caption{Energy calibration of the fit parameter $a$: The calibration results from 1064~nm single photons. 
  The standard deviation $\sigma$ and the corresponding regions result from the PSA of the representative signal sample. 
}
{\begin{tabular}{cccc}
    \hline
  		& $a$ & $E$ [eV]	& $\lambda$ [nm] 	\\
    \hline
    mean $\mu$ & 1.01 & 1.165 & 1064  \\
    (full) 3$\sigma$-region & 0.82-1.20 & 0.948-1.382 & 1308-897  \\
    half (high-energetic) 3$\sigma$-region & 1.01-1.20 & 1.165-1.382 & 1064-897  \\
    \hline
\end{tabular}}
\label{tab:fit_calibration}
\end{table}

\section{Background events and dark count rate}
\label{sec:background}

Because an ALPS detector is required to have low dark count rates (Sec.~\ref{sec:intro}), we extensively studied the background of the ALPS TES detector.
Using different fiber setups, we undertook several long-term measurements (at least $>15$~h), and analyzed the trigger events by applying PSA. Dark counts are defined as events which lie in the signal region. A dark count rate, $DC$, is calculated by dividing the total counts by the total measurement time.
{In the following, we mainly present two measurements and analyses:
  \begin{itemize}
    \item Using no fiber and a dark TES in order to analyze the intrinsic background (Sec.~\ref{sec:background:intrinsic}).
    \item Using a fiber-coupled TES in order to analyze the thermal photonic background resulting from the warm fiber end (Sec.~\ref{sec:background:thermal}). 
  \end{itemize}
}

\subsection{Intrinsic background}
\label{sec:background:intrinsic}

Firstly, we investigated possible noise events of the read-out system \cite{jde2014}. Therefore, we operated only the SQUID with a normal resistive TES. We observed few events with a short-time baseline shift. We interpret these events as environmental electromagnetic interference.
For example, switching on and off the laboratory lights or maybe starting a machine adjacent to the laboratory can cause such baseline shifts.
All these events are removed by PSA and do not result in dark counts, however, they may lead to an increase of the dead time fraction of $<0.02$~\%, defined as the ratio of total accumulated dead time and measurement time.

\begin{figure}
\begin{center}
\begin{minipage}{70mm}
\resizebox*{7cm}{!}{\includegraphics{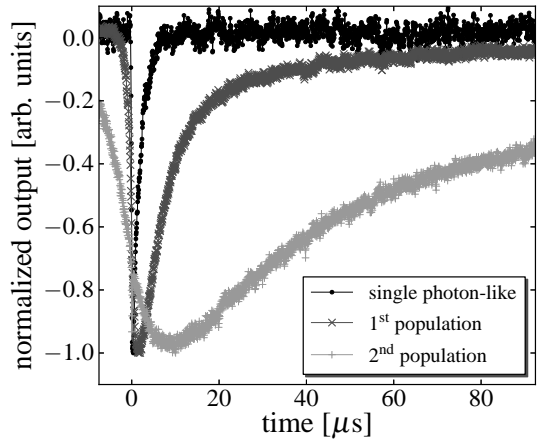}}
\caption{
{Intrinsic background: The plot} shows the normalized and averaged pulse of different populations: a single-photon like population and two background populations with longer time constants.
}
\label{fig:intrinsic} 
\end{minipage}
\end{center}
\end{figure}

Secondly, we analyzed the intrinsic background of the ALPS detector \cite{jde2014}. 
Therefore, we operated a ``dark'' TES chip without a fiber, and the chip surface was facing the 80~mK shield. 
We determined three populations of calorimetric pulses recorded with this setup (Fig.~\ref{fig:intrinsic}). 
Besides single photon-like events, we have observed two background populations with a greater value of the exponential falling time constant:
\begin{itemize}
  \item single photon-like events with $\tau_{\rm eff} = 1.53$~$\mu$s (Sec.~\ref{sec:alps_tes_detector:tes}) and a rate of $\sim 1 \cdot 10^{-4}$~s$^{-1}$ (see below)
  \item 1$^{\rm st}$ background population with $\tau_1 \approx 13.1$~$\mu$s and a rate of $\sim 5 \cdot 10^{-3}$~s$^{-1}$
  \item 2$^{\rm nd}$ background population with $\tau_2 \approx 73.0$~$\mu$s and a rate of $\sim 5 \cdot 10^{-3}$~s$^{-1}$
\end{itemize}

The origin of these two background populations is most likely related to radioactivity and cosmic muons. 
Firstly, any charged particle could ionize silicon layers of the TES chips and cause a heat up. 
The two populations could be related to two different types of silicon layers around the tungsten film. This could explain two different thermal time constants. 
Secondly, both populations show a uniform broad energy spectrum. 
Thirdly, an estimate of the event rate fits roughly the measured rate.
Fourthly, we observed several coincidence events when operating both TES channels, which are located close together.
Such background events are discriminated by the PSA method and do not contribute to a dark count rate. 
The dead time contribution of these background events is $<0.001$~\% and {thus} negligible.

The photon-like events could be photons occurring during an ionization process  of the sensor chip.
A second possible explanation are thermal photons from warmer regions inside the cryostat. 
We observe that the thermal shields inside the ADR are not totally light-tight. 
In addition, the spectrum of photon-like events is not uniform  and shows more low-energetic events, which could point to a black body spectrum (Sec.~\ref{sec:background:thermal}).

Applying the PSA to four long-term measurements, we determined 22 dark counts \cite{jde2014}, which lie within the 3$\sigma$-region of 1064~nm photon signals (Tab.~\ref{tab:fit_calibration}). 
This results in an intrinsic dark count rate of
$$DC_{\rm intrinsic} = (1.0 \pm 0.2) \cdot 10^{-4} \,\, {\rm s}^{-1}.$$
We consider this value as a fundamental limit of dark noise when detecting 1064~nm signals with this TES detector setup.
Further reduction is possible after improving the light-tightness of the inner ADR and optimizing the design to avoid photons resulting from an ionization of the chip substrate.

\subsection{Thermal photonic background}
\label{sec:background:thermal}

The dark count rate of a fiber-coupled TES, $DC_{\rm fiber}$, is necessary for an application.
Here, the ``warm'' fiber end is located outside the cryostat in order to connect it to an optical experiment. 
It has been observed by other groups that photons from room-temperature surfaces represent a significant background \cite{cabrera1998}. 
The spectrum of thermal photons is in accordance with a room-temperature (300~K) black body spectrum folded by the fiber transmittance.
For example, there is a cutoff for wavelengths $> 2000$~nm for silica fibers.
However, other groups concentrated mainly on the telecommunication wavelength of 1550~nm as a signal \cite{rosenberg2005, miller2007, fujii2011}. 
A dark count rate for 1064~nm signals has to our knowledge not been published.

To determine $DC_{\rm fiber}$ for 1064~nm signals, we performed several long-term measurements \cite{jde2014}. 
The total fiber was 2.2~m long and its end, outside of the cryostat, was covered to suppress stray light.  
Due to the limited {read-out} bandwidth, we had to adjust the trigger level to reduce the trigger rate,
otherwise measurements would have been affected by dead time due to a high rate of low-energetic thermal photons. 
We set the trigger to a level which corresponds to a value of nearly $a = 1$. In the following, we consider the high-energetic half of the 3$\sigma$-region (Tab.~\ref{tab:fit_calibration}). 

Applying the PSA to two representative long-term measurements, we determined that \cite{jde2014}:
\begin{eqnarray}
DC_{\rm fiber} = (8.6 \pm 1.1) \cdot 10^{-3}\,\,{\rm s}^{-1} \ .
\label{eqn:dc:fiber}
 \end{eqnarray}
This value corresponds to the rate for photons with a wavelength going from 897 to 1064~nm including the detector's energy resolution (Eqn.~\ref{eqn:erg_res:psa}). 
Furthermore, it corresponds to an overall detection efficiency of $\sim23$~\% for 1064~nm signals. This includes optical losses, the chosen trigger level and  the efficiency of the analysis.%
\footnote{
The optical losses are mainly caused by the used fiber setup. 
Using a commercial power meter, we measured a transmission of only $\sim60$~\% due to two fibers with a fiber connector in between.
In addition, we conservatively estimate a 95~\%-efficiency of the fiber-to-TES coupling (Sec.~\ref{sec:alps_tes_detector:tes}).
Because of the chosen trigger level, only $\sim 80$~\% of the nominal signals are considered. 
Furthermore, the two-dimensional Gaussian 3$\sigma$-region contains 98.9~\% nominally.
Finally, the half 3$\sigma$-region includes a factor 0.5.}

\begin{figure}
\begin{center}
\begin{minipage}{70mm}
\resizebox*{7cm}{!}{\includegraphics{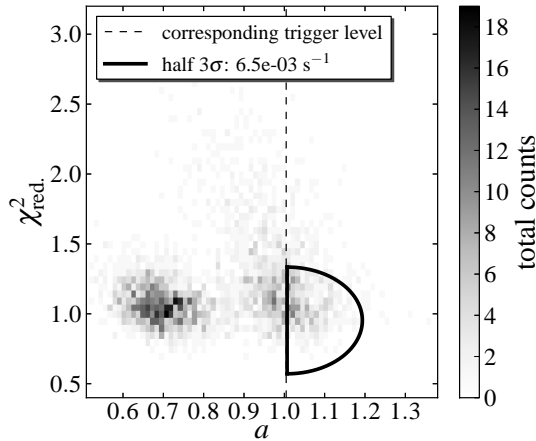}}
\caption{
{Thermal photonic background: It is shown} a typical result of a long-term measurement analyzed with PSA. 
  We only consider the high-energetic half of the 3$\sigma$-region due to the adjusted trigger level. 
}
\label{fig:fiber} 
\end{minipage}

\end{center}
\end{figure}

In Fig.~\ref{fig:fiber}, the distribution of events, the signal region and the corresponding trigger level are depicted. Two photon-like populations are identified:
\begin{itemize}
  \item \textbf{``photon-baseline noise'' pile-up events:} 
This population is located around $a=0.7$ with $\chi^2_{\rm red.} < 1.5$.
These events are related to low-energetic photons which can reach the trigger level because of a coincidental noise contribution near the peak height. 
  
  \item \textbf{``photon-photon'' pile-up events:}
This population has about $a=1.0$ with a range of $\chi^2_{\rm red.} = 0.7 \ldots 3.0$.
By considering an event with 
$\chi^2_{\rm red.}> \mu_{\chi^2_{\rm red.}} + 3 \sigma_{\chi^2_{\rm red.}} = 1.34$,
it clearly appears as a first-order pile-up event of two photonic constituents.
By considering events with
$\chi^2_{\rm red.} \approx \mu_{\chi^2_{\rm red.}} + 3 \sigma_{\chi^2_{\rm red.}} = 1.34$,
we can estimate a time resolution of the PSA method, $\tau_{\rm PSA} \approx 0.5$~$\mu$s.
If two photon-like events occur within $\tau_{\rm PSA}$, this first-order pile-up event cannot be discriminated by the PSA and is counted as a single photon-like event and, in case of matching energy, as a dark count of 1064~nm signals. 
\end{itemize}
It seems that the half 3$\sigma$-region is dominated by {``photon-photon''} pile-up events {(Fig.~\ref{fig:fiber})}.
This assumption is discussed in the following paragraph. 
We compare the measured rate to an expected rate using
\begin{enumerate}
  \item a model with a black body spectrum  and  a detector with a perfect time resolution (idealized), 
  \item a measured thermal photon rate with a limited time resolution (realistic).
\end{enumerate}

First, for an idealized model, we assume a 300~K black body spectrum, a single-photon detector with 10~\% energy resolution, perfect time resolution, no optical losses and overestimated parameters for the fiber core area and aperture.
This results in $DC_{\rm ideal.~model} = 3.4 \cdot 10^{-4}$~s$^{-1}$ \cite{jde2014}. 
Thus, the measured rate (Eqn.~\ref{eqn:dc:fiber}) is $\sim1.5$ orders of magnitude higher than $DC_{\rm ideal.~model}$.
This suggests that the measured dark count rate cannot be dominated by single thermal photons. 

Second, we estimate an effective rate, $\dot{n}_{\rm eff}$, caused by first-order pile-ups by considering accidental coincidences. It is \cite{eckart1938}
\begin{eqnarray}
  \dot{n}_{\rm eff} = 2\, \tau \, \dot{n}_1 \, \dot{n}_2 \ ,
\label{eqn:acc_coin}
\end{eqnarray}
where $\tau$ is the resolving time and $\dot{n}_1$ and $\dot{n}_2$ uncorrelated rates. We set $\tau = \tau_{\rm PSA} \approx 0.5$~$\mu$s as current time resolution of the ALPS TES detector. 
{We estimate $\dot{n}_1 \approx \dot{n}_2 \approx 10^{2}$~s$^{-1}$ from a pulse height measurement {(Fig.~4.3b in \cite{jde2014})}.}
Here, $\dot{n}_1$ contains thermal photons with wavelengths going from 1600 to 1900~nm and $\dot{n}_2$ from approximately 1900 to 2200~nm. 
Photons of these two contributions can effectively combine to single photon-like events lying in the half 3$\sigma$-region of 1064~nm signals.
It is $\dot{n}_{\rm eff} \simeq 10^{-2}$~s$^{-1}$, which is in the same order of magnitude as the observed dark count rate in Eqn.~\ref{eqn:dc:fiber}.

 
\section{Summary}
\label{sec:concl}

The NIST-fabricated TES has been successfully operated in an ADR. 
Routine operations in the ALPS lab environment have been carried out with measurements on the performance of the system including stability, characteristics of the TES sensor and its readout as well as background measurements.
A pulse-shape analysis has been developed using a self-consistent pulse template to fit individual pulses. 
Energy reconstruction and background rejection have been demonstrated to work well (relative energy resolution of $\Delta E/E\approx 6$~\% and background rejection removing spurious events from ionization as well as electromagnetic induced events). 

The overall detection efficiency of $\sim 23~\%$ with a background rate of $\sim 9$~mHz  (Eqn.~\ref{eqn:dc:fiber}) from a warm fiber has been achieved with this first setup. 
The main requirements of the ALPS setup (single-photon detection efficiency $>70$~\% and background rate in the mHz regime) are achievable with an improved setup. 
The reduction of thermal background is possible with a fiber cutting off at a smaller wavelength. 
This will relax the required event selection to achieve an efficiency as required for the ALPS experiment.

 
\section*{Acknowledgements}
We want to thank NIST, Boulder, US, for the TES devices and PTB, Berlin, Germany, for the SQUID sensors. We also want to thank our ALPS collaborators, especially S.~Ghazaryan, E.~von Seggern and C.~Weinsheimer.



\end{document}